# Revisiting Wireless Internet Connectivity: 5G vs Wi-Fi 6

Edward J Oughton[1*], William Lehr[2], Konstantinos Katsaros[3], Ioannis Selinis[4], Dean Bubley[5], Julius Kusuma[6]

[1]George Mason University, Fairfax, VA
[2]Massachusetts Institute of Technology, Cambridge, MA
[3]Digital Catapult, London, UK
[4]5G Innovation Centre, University of Surrey, Guildford, UK
[5]Disruptive Analysis, London, UK
[6]Facebook Connectivity, Menlo Park, CA

*Corresponding author: Edward J. Oughton (E-mail: eoughton@gmu.edu; Address: GGS, George Mason University, 4400 University Drive, Fairfax, VA)

## Abstract

In recent years, significant attention has been directed toward the fifth generation of wireless broadband connectivity known as '5G', currently being deployed by Mobile Network Operators. Surprisingly, there has been considerably less attention paid to 'Wi-Fi 6', the new IEEE 802.1ax standard in the family of Wireless Local Area Network technologies with features targeting private, edge-networks. This paper revisits the suitability of cellular and Wi-Fi in delivering high-speed wireless Internet connectivity. Both technologies aspire to deliver significantly enhanced performance, enabling each to deliver much faster wireless broadband connectivity, and provide further support for the Internet of Things and Machine-to-Machine communications, positioning the two technologies as technical substitutes in many usage scenarios. We conclude that both are likely to play important roles in the future, and simultaneously serve as competitors and complements. We anticipate that 5G will remain the preferred technology for wide-area coverage, while Wi-Fi 6 will remain the preferred technology for indoor use, thanks to its much lower deployment costs. However, the traditional boundaries that differentiated earlier generations of cellular and Wi-Fi are blurring. Proponents of one technology may argue for the benefits of their chosen technology displacing the other, requesting regulatory policies that would serve to tilt the marketplace in their favour. We believe such efforts need to be resisted, and that both technologies have important roles to play in the marketplace, based on the needs of heterogeneous use cases. Both technologies should contribute to achieving the goal of providing affordable, reliable, and ubiquitously available high-capacity wireless broadband connectivity.

## 1. Introduction

Almost in synchrony we are seeing the roll-out of the next generation of wireless technologies for both cellular and Wi-Fi connectivity. While there has been much excitement around the world regarding the fifth generation of cellular technology known as '5G', there is comparable enthusiasm for the next version of the Institute of Electrical and Electronics Engineers' (IEEE) 802.11 Wireless Local Access Network (WLAN) standard, 'Wi-Fi 6'. Next generation wireless connectivity technologies are needed to further enable the shift to a Digital Economy given the productivity and social benefits that a successful transition promises (Bauer, 2018; Graham and Dutton, 2019; Hall et al., 2016a, 2016b; Mansell, 1999; Parker et al., 2014).

Competition between cellular and Wi-Fi is not a new debate. Two decades ago, 3G and an earlier Wi-Fi 801.11 standard were seen as competing wireless technologies (Lehr and McKnight, 2003). This was before Apple's iPhone propelled mass-market mobile broadband via smartphone devices toward becoming a must-have platform for ubiquitous Internet connectivity (West and Mace, 2010). In that earlier time, the question was whether 3G and Wi-Fi would be competitors or complements. Upon reflection, the two technologies were predominantly complementary, with Wi-Fi providing high-capacity indoor hotspots for broadband, and cellular providing connectivity outdoors and to support high-speed mobile access. The widespread adoption of Wi-Fi in portable consumer devices such as laptops and the original iPhone[1] fuelled demand growth for mobile computing and data access which later helped drive rapid adoption of 4G mobile broadband services.

---

[1] For example, the first iPhone launched did not have 3G connectivity, instead relying on a mix of 2G GSM and Wi-Fi connectivity.

The earlier analysis by Lehr & McKnight (2003) focused on the roles these technologies played in providing consumer broadband services, rather than for wireless connectivity by business enterprises, which has always involved a more complex array of wireless technologies. Businesses were earlier and heavier users of computing and data communication services than consumers and were the first to deploy WLAN technologies. Cellular has historically focused on the mass-market and has played less of a role in providing wireless connectivity for businesses, except where the two worlds overlap.[2] This is changing since many anticipate that some of the most interesting and important applications for 5G will be in vertical industrial sectors and other private-network applications (which are often also indoors or deployed in campus environments and hence are less dependent on the licensed spectrum that cellular operators have principally relied on in the past).

Given that we are in the early stages of the transition to the next generation of cellular and WLAN technologies, with the potential to anticipate the significant wireless connectivity changes that might occur, the research question we explore in this paper is as follows:

> *To what extent will 5G and Wi-Fi 6 be predominantly complementary, or will technological substitution lead to a new trajectory for wireless connectivity, with one gaining increased prominence over the other?*

Although the debate among technology and industry partisans favouring cellular or Wi-Fi technology has been heated, the research community has not adequately considered the extent to which these technologies may interact both as competing alternatives in some contexts and as complementary technologies in others. For example, the introduction of a 5G standard (5G NR-U) which can operate in the unlicensed spectrum bands (traditionally the domain of Wi-Fi) led Qualcomm to hypothesise that 5G might result in the demise of Wi-Fi (Light Reading, 2019) and some proponents of cellular have made similar bold claims that 5G will 'kill-off' Wi-Fi 6 (Bloomberg, 2017).

---

[2] Examples include consumers who are also employees needing wide-area access or the businesses that by nature require access to ubiquitous wireless (e.g., transport services).

In a 3G/4G world, many cellular users have preferred to access data intensive applications via Wi-Fi because of better performance, and to avoid incurring the higher costs or data caps associated with cellular services. With enhanced performance of 5G and with the shift to unlimited data plans for mobile, some expect end-users to shift their data traffic from Wi-Fi to cellular networks. Additionally, the standards body 3GPP has now included indoor broadband scenarios proposing the use of cellular technologies to deliver connectivity in office settings, hence raising the potential for 5G to compete directly with a key Wi-Fi use case (3GPP, 2016a).

In this paper, we compare and contrast alternative perspectives on the relative merits of 5G and Wi-Fi 6, as discussed in the technical, trade press, and academic literature and drawing on our collective expertise in advising industry stakeholders and policymakers around the world on the implications of alternative wireless technologies over the past several decades. We begin in Section 2 by reviewing the demand-side changes which will affect wireless Internet connectivity over the next decade. In Sections 3 and 4, we provide a general qualitative overview of 5G and Wi-Fi 6 for a policy and economics audience, and then in Section 5, we compare and contrast these technologies. Finally, Section 6 offers concluding thoughts.

## 2. Future demand-side changes affecting wireless connectivity

Global data flows have been rapidly growing for decades, with the majority consisting of Internet Protocol (IP) traffic carried over the open Internet and/or private IP networks (Claffy et al., 2020; Knieps and

Page 4 of 42

Stocker, 2019). For example, over the next decade two thirds of the global population will be online (>5.3 billion), and the number of Internet-connected devices will exceed more than three times the global population (>29 billion). While the traffic growth experienced through the 1990s and 2000s was predominantly due to increased penetration of fixed broadband Internet services, this is now being driven by growth in wireless broadband. Most of the generated traffic will be served by Wi-Fi, with about one fifth being served by cellular (Cisco, 2020). There has generally been a co-evolving relationship between greater availability of wireless Internet connectivity fuelled by the proliferation of Wi-Fi hotspots, and the expanding coverage and improved performance of 4G. When combined with much more capable devices there has been growing demand for near-ubiquitous connectivity to broadband content, applications and services (Stocker et al., 2017).

In this review we identify a variety of demand-side factors which will affect the future of wireless Internet connectivity, which will be driven by the increasing number, and changing composition, of devices, the ongoing rise in the quantity of data generated per device, and the growing number of total users (which increasingly may include 'things').

Since the original debate regarding wireless broadband began between 3G and Wi-Fi almost twenty years ago, consumer device ownership patterns have shifted considerably. Figure 1 (A) illustrates how device adoption has evolved in the United States, which demonstrates a pattern that has been echoed in Europe and in several other high-income markets.[3]

The exhibit highlights multiple important trends: (1) a shift from fixed landlines to cellular phones, leading first to the loss in second lines, and then to a growing number of households becoming cellular-only;

---

[3] See UK (Ofcom, 2020) or European Union (Eurostat, 2020).

(2) increased take-up of data-capable/Internet-connectable smartphones, connected first via Wi-Fi for broadband but increasingly with 4G after roll-out began in 2010; (3) expanded adoption of other wirelessly-connected devices like tablets, e-readers, etc., supplementing personal computer connectivity options; and (4) increasing numbers of Internet-connected devices per user (as more users have multiple devices).

*Figure 1 (A) Technology Trends (US) (Roser et al., 2019) and (B) Global IoT/M2M connections by use case (Cisco, 2020)*

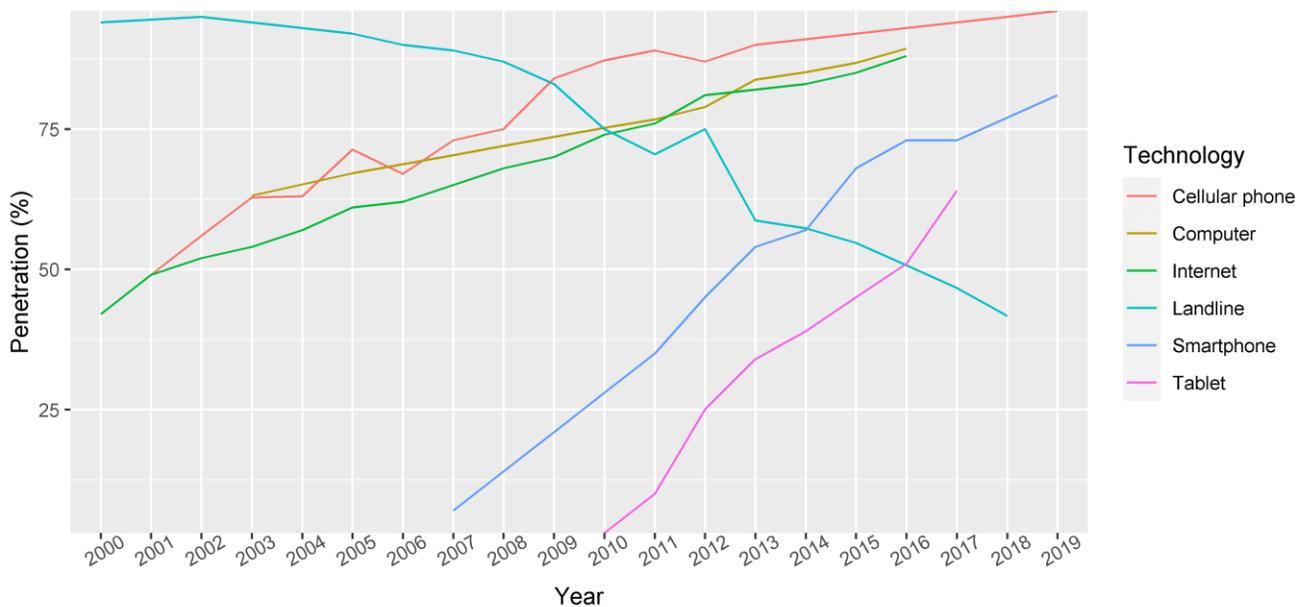

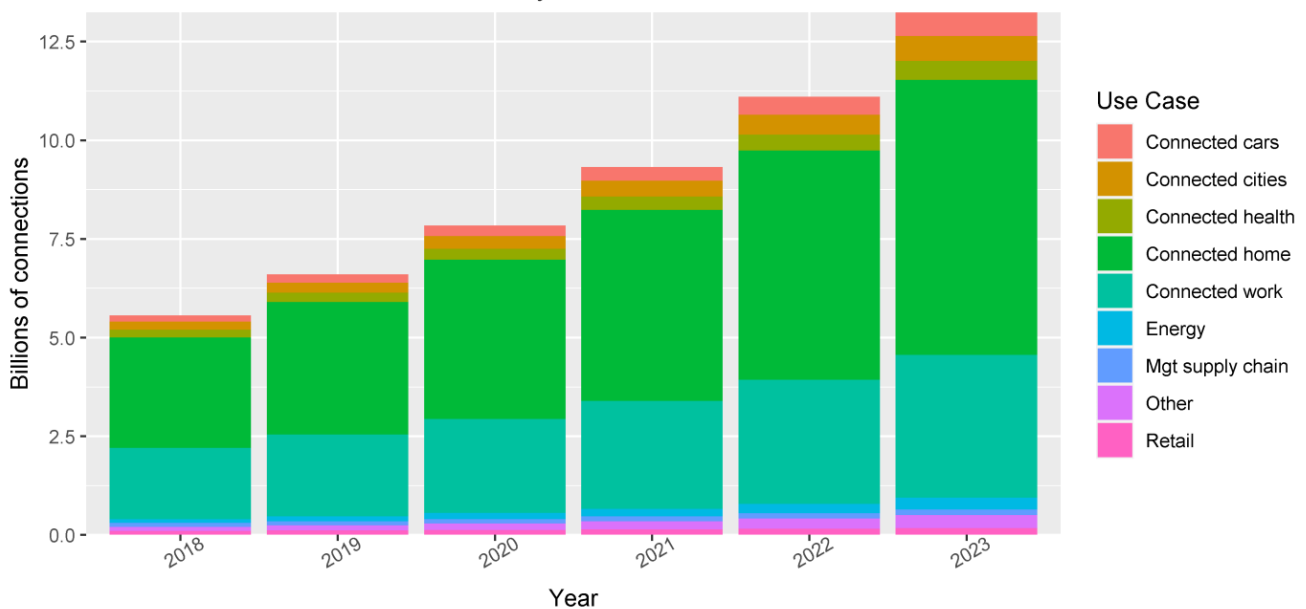

(A) Interpolated values for Landlines (2006-2007), Computers (2004, 2006-2009) and Smartphones (2007-2009)

Figure 1 (B) illustrates how the largest number of global connections are in the home or office, and thus more likely to use Wi-Fi as a form of wireless Internet connectivity. Additionally, there are a number of key trends which include: (1) M2M and IoT devices being the primary driver for the increasing number of connections, moving from approximately 8 billion globally in 2020 to over 14 billion by 2023; (2) the two major use cases involve the connected home and connected workplace; and (3) the other M2M and IoT use cases will be relatively minor in comparison, including connected health, cities and cars.

Finally, the ongoing improvements in video quality will continue to increase the quantity of traffic generated per device. For example, huge advances in network capacity have enabled the growth of higher-data-rate applications such as High Definition (HD) video conferencing, streaming entertainment media, and highly interactive gaming replacing lower-data rate text and voice-only communication services. Currently video accounts for over three quarters of total consumer and household traffic, and has a multiplier effect whereby an Internet-connected HD TV generates as much daily traffic as an average household (Cisco, 2020). With the ongoing shift to higher quality video there is an even greater requirement for increased connection capacity, which has an impact on wireless broadband requirements. For example, Netflix is one of the most popular video streaming platforms and can provide a service on relatively low connection capacity, ranging from 0.5-1.5 Mbps (Netflix, 2020). However, Standard Definition (SD) video requires at least 3 Mbps, HD requires at least 5 Mbps and Ultra High Definition (UHD) requires at least 25 Mbps. Thus, as consumer preferences increasingly move towards a minimum of HD, but preferably UHD video quality, the quantity of data demand generated per device will increase, which affects demand for wireless connectivity. Video is expected to continue driving global consumer data demand over the next decade, resulting from more devices serving users with better quality streamed content.

Additional use cases which may drive demand over the next decade may relate less to the quantity of traffic generated, and depend more on the promised Quality of Service that a wireless service can guarantee. First, a significant issue with 4G mobile broadband is latency. Indeed, attempting to use certain services such as real-time video (e.g. to support Augmented Reality or Virtual Reality) may be badly affected by latency response times on the order of 100 ms. Secondly, and related to the first, variable reliability when using wide-area mobile connectivity can severely affect the user experience for more demanding latency-sensitive services.

## 3. An overview of 5G technical features

The development body responsible for cellular standardisation efforts is the 3rd Generation Partnership Project (3GPP), which is the industry organisation that defines the global specifications for 3G, 4G and 5G technologies.[4] For 5G, the first specification released by 3GPP for Phase 1 (Release 15) states there are three key technical use cases (3GPP, 2019), including

1. Enhanced Mobile Broadband (eMBB)
2. Ultra Reliable and Low Latency Communications (URLLC)
3. Massive Machine Type Communications (mMTC)

Within 5G eMBB, one of the first use cases is using this approach to provide broadband via Fixed Wireless connectivity. Additionally, URLLC is technically made up of multiple use cases, either Ultra

---

[4] 3GPP is an umbrella term for seven individual organisational partners from Asia, Europe and North America which collaboratively set cellular standards. For more information on 3GPP, see https://www.3gpp.org/about-3gpp.

Reliable or Low Latency communications, or a combination thereof. The launch of Release 16 takes place in 2020, followed by Release 17 in 2022, following an approximate 15-month standardisation (3GPP, 2020).

The aim of eMBB is to move beyond what is capable in 4G to provide improved data-rates, traffic/connection density and user mobility (Cave, 2018; Oughton et al., 2019; Oughton and Frias, 2016). The use case is expected to be delivered for a range of coverage scenarios and applications such as streaming, video conferencing, Augmented Reality (AR) and Virtual Reality (VR). This includes different service areas ranging from indoor to outdoor, urban to rural, home to office, and local to wide area connectivity, as well as special deployment circumstances for mass gatherings, broadcasting, residential access and vehicles travelling at very high speeds. Ultrafast mobile broadband of 100 Mbps is expected outdoor (for the mean user experienced throughput), with peak throughput up to 10 Gbps on an indoor 5G network (3GPP, 2016b), should sufficient spectrum be available and network conditions allow.

In contrast, the aim of delivering low latency and highly reliable communication services is driven by new industrial automation applications in vertical sectors (manufacturing, automotive etc.) (Vuojala et al., 2019). Current 4G systems can experience significant latency issues resulting from delay on the radio interface, transmission within the system, transmission to a server which may be outside the system, and data processing. Hence, 5G aims to reduce this latency through the RAN and core, along with taking advantage of local service hosting called 'edge computing'. Note though, that edge computing is not limited to 5G networks, as it can support multi-access networks, including both cellular and Wi-Fi. However, the architecture of 5G, with the increased use of virtualisation of network infrastructure, compared to previous cellular generations and Wi-Fi, makes better use of edge computing capabilities. The aim is to provide reliability of 99.9999% for process automatic, with a data rate <100 Mbps and an end-to-end latency of <1-2 ms for the user plane and less than 10 ms for the control plane (3GPP, 2016c).

Additionally, mMTC is extending LTE IoT capabilities—for example, through 4G-based narrowband IoT (NB-IoT) to support huge numbers of devices with lower costs, enhanced coverage, and long battery life, reaching thousands of end-devices (3GPP, 2016d). Later 3GPP releases (e.g. 17 or 18) are expected to provide a narrowband IoT capability using the 5G New Radio interface (5G PPP Architecture Working Group, 2019).

Taking advantage of the proposed architectural structure, one of the fundamental features that is being supported is infrastructure 'slicing'. Network slicing requires a continuous adjustment of customer-centric Service Level Agreements (SLAs) with infrastructure-level network performance capabilities. As customers such as vertical industries request new types of connectivity services from providers, both the creation and operation of these services will have to demonstrate a very high level of automation to ensure very efficient lifecycle management of network slice instances, via the use of an end-to-end framework.

The proposed 5G architecture is accomplished in a recursive structure, which can be specified as a procedure that is applied repeatedly. This philosophy increases scalability since the same service category can be deployed repeatedly, and simultaneously, at different locations. From the perspective of virtualised infrastructure, this recursive approach permits the operation of a slice instance on top of the resources provisioned by another slice instance. As an example, each tenant can own and deploy its own MANagement and Orchestration (MANO) system using these principles.

Several of the more important technological capabilities in 5G include the introduction of support for (1) millimetre-wave frequencies, (2) Massive Multiple Input Multiple Output (mMIMO), (3) increased use of small cells, (4) advanced coding and modulation, (5) splitting of the control/user plane, (6) beamforming and (7) full duplex data transfer. Each of these developments expands the capability of 5G to support higher data rates, enhanced spectral efficiency, and greater reliability in the face of variable spectral environments.

For example, the millimetre-wave band (technically 30–300 GHz, although commonly all bands above 20GHz are called 'mmWave') contains over 90% of the allocated radio spectrum, and yet most of the mmWave spectrum is significantly under-utilised (Niu et al., 2015; Roh et al., 2014). Previous generations of cellular technology made little use of these frequencies due to its poor propagation characteristics and the limitations imposed by then-available digital technologies. Advances in wireless and networking technologies (including those noted above), the continued growth in demand for wireless connectivity offering ever-faster data rates, along with the facts that this spectrum is under-utilised (and hence less congested) and wider-frequency channels are available (then at frequencies below 20GHz) makes this higher frequency spectrum attractive. Empirical evidence already demonstrates that MNOs preference frequencies with large bandwidths for deployment (Frias et al., 2020). Early results indicate that significant Non Line Of Sight (NLOS) outdoor street-level coverage using mmWave is achievable within approximately 200 meters of the serving cell (Akdeniz et al., 2014; Rangan et al., 2014).[5]

Even without the propagation challenges raised by using higher-frequency spectrum, MNOs are shifting to small(er) cell architectures to provide the capacity needed to support traffic growth in hot-spot areas. This requires a massive densification in the number of cells required to serve their coverage area footprints, which more the largest MNOs, are national. The shift to small cells facilitates spatial reuse (expanding the capacity of scarce spectrum resources[6]), provides better support for lower-power devices (since the distance over-which signals need to travel is shorter and hence less power is required), and makes it feasible to access a wider array of spectral resources, including in the mmWave bands. While poor network planning always leads to significant cost ramifications (Haddaji et al., 2018; Taufique et al., 2017; Wisely et al., 2018; Yaghoubi et al., 2018), the mass deployment of small cells presents a set of unique challenges related to spectrum management, energy efficiency and the logistics of deploying backhaul (Ge et al., 2016, 2014; Wang et al., 2015). One way to address some of these issues is the disaggregation of small cell architectures, based on the virtualisation principles discussed later in relation to Figure 2.

---

[5] Using mmWave spectrum to build out 5G networks based will result in several important changes in the design and performance of mobile broadband networks, including (i) a large increase in the number of antennas, (ii) propagated signals being more sensitive to blockages, (iii) variable propagation laws (with NLOS being far worse than LOS) and (iv) fewer multipath components in the radio channel being used (Bai et al., 2014). One way MNOs will respond to these challenges is by ensuring that their networks can make use of a portfolio of spectral resources so they are not solely reliant on mmWave spectrum. Other technologies like MIMO and beam-forming antennas also will contribute to making 5G small cells able to address the challenges that arise from the need to support more QoS demanding wireless applications (e.g., VR/AR) in heterogeneous networking environments.

The deployment of mMIMO technologies are a key capacity enhancing technique (Bogale and Le, 2016; Jungnickel et al., 2014; Mumtaz et al., 2016). Compared to 4G MIMO, 5G mMIMO provides large spectral efficiency gains by taking advantage of multiple antennas at each Base Station (BS).[7] However, integrating mMIMO into future 5G networks will confront the challenges of higher equipment costs and energy consumption associated with delivering the more demanding services utilising higher frequency spectrum (Panzner et al., 2014), (Huang et al., 2017).

---

[6] In scenarios where small cells are deployed within macro cell areas there has been shown to be a beneficial effect in spectral efficiency (Jungnickel et al., 2014)).

[7] The multiple antennas facilitate better beam-forming, by which the transmitted signals may be better focused through spatial filtering at the transmitter (precoding) and/or the receiver (Papadopoulos et al., 2016), leading to improvements in the received Signal to Interference Plus Noise Ratio (SINR). Multiplexing gains can also be achieved, with multiple streams of information being transmitted simultaneously. By adding additional antennas at each site additional overheads in Channel State Information (CSI) are avoided, focusing the radiated energy toward the intended directions while minimising intra-and intercell interference (Boccardi et al., 2014).

The use of beamforming in 5G is a key feature and enables a single stream of information to be focused towards a user, rather than being transmitted radially, significantly reducing the level of interference experienced at other cells. In combination with mMIMO, a BS can then estimate the most efficient route to send information packages based on reduced interference, by triangulating the location of the user device.

Basic radio antennas can only perform a single task at one time, such as transmitting or receiving information. In 5G, the introduction of full duplex model allows both UL and DL directions on a single stream. So rather than using Frequency Division Duplexing (FDD) where streams are split into UL and DL channels, or Time Division Duplexing (TDD) where information travels in just one direction at one point in time, bidirectional signals may be sent simultaneously in full duplex with 5G.

To deliver the technical specifications of the 5G standard in a cost-efficient way MNOs are examining the migration of a traditionally Distributed RAN (D-RAN) characterised by the co-location of Base Band Units (BBUs) and Remote Radio Heads (RRHs), to a Centralised/Cloud-RAN, as illustrated at the top of Figure 2. A C-RAN architecture would consist of a central location providing shared BBU resources to reduce capital expenditure, operational expenditure and ultimately the Total Cost of Ownership, with RRHs connected directly to the pool of BBUs via high bandwidth, low latency transport links known as fronthaul, as illustrated at the bottom of Figure *2*. As a part of the 3GPP framework, multiple functional splitting options, one of which covers C-RAN (Option 8), have been proposed to meet the diverse requirements of 5G (3GPP, 2016d). OpenRAN has introduced a set of open API specifications between the components comprising this disaggregated RAN, namely the Central Unit (CU), Distributed Unit (DU) and Radio Unit (RU). The transition to a virtualised RAN architecture supports the disaggregation of both hardware and software components, facilitating the use of general-purpose hardware and software. This general-purpose platform architecture reduces total costs relative to relying on specialised hardware or software and allows the general-purpose hardware/software components to be utilised more efficiently, contributing to the realisation of sharing economies. Moreover, this allows for flexibility and adaptability in the implementation and how network resources may be sliced (virtualised). Meeting the need for high bandwidth, very low latency links across the fabric of the dense small cell edge RAN networks requires a mix of innovative back-haul, mid-haul and front-haul solutions. These include Integrated Access and Backhaul (IAB), Hybrid Fibre Coax (HFC), and Nexgen Fibre. A recent analysis found a positive 5G business case for eMBB over the period 2020-2030 (Rendon Schneir et al., 2019).

*Figure 2 The evolution of cellular RAN configurations (Alsharif and Nordin, 2017)*

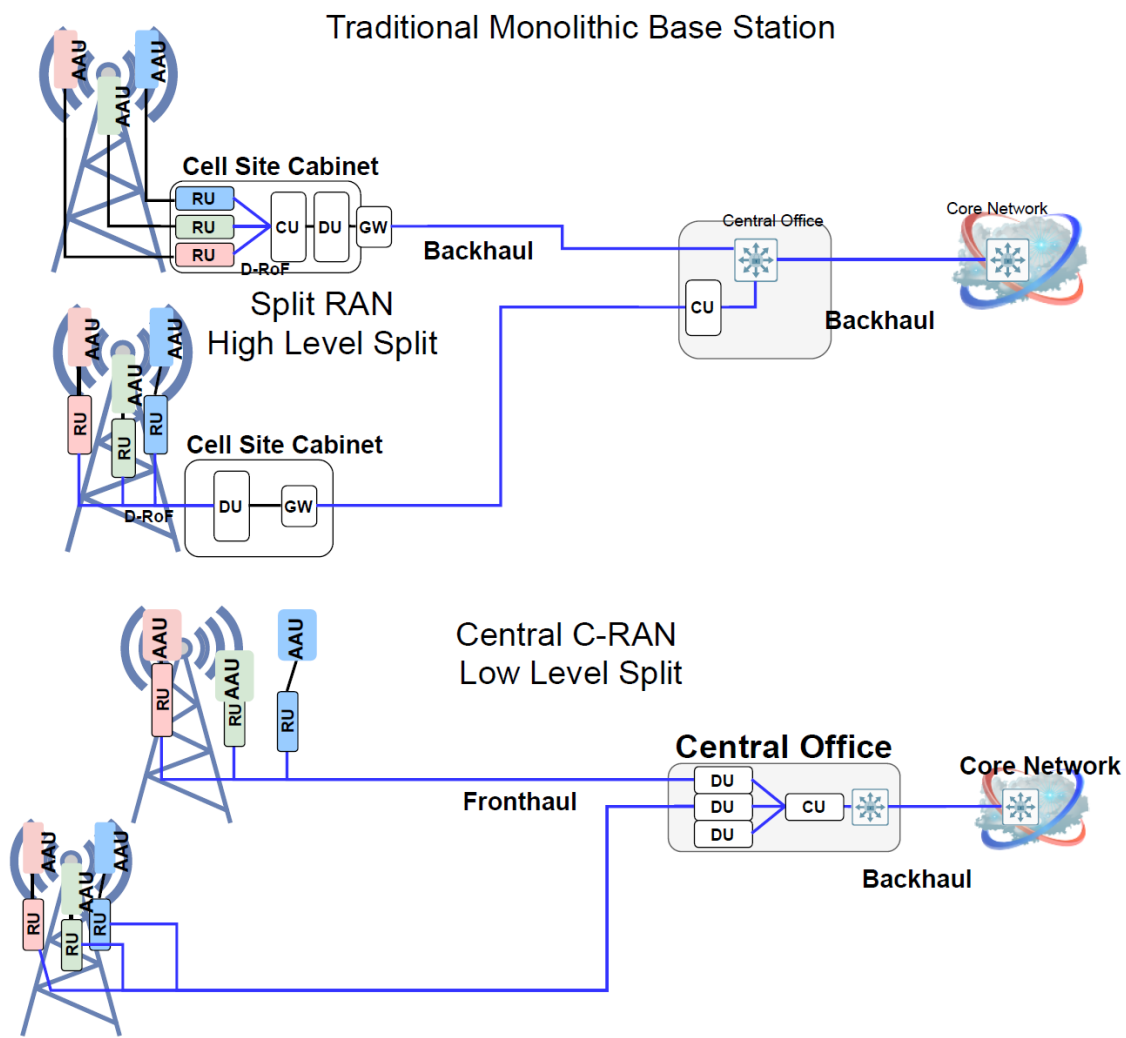

## 4. An overview of Wi-Fi 6 (802.11ax) technical features

In this section, we turn to highlighting the main features of the Wi-Fi 6 technology. Two different industry organisations have led in the development of Wi-Fi. First, the IEEE's Project 802 is the development body responsible for many networking standards, including the suite of Wi-Fi technologies. Second, the Wi-Fi Alliance is a non-profit organisation comprised of a global network of companies tasked with ensuring interoperability and certifying and promoting different Wi-Fi products (including adding more accessible marketing terms such as the Wi-Fi 4, 5 or 6 labels).[8] This includes being responsible for both technical aspects, such as creating additional specifications for products such as Wi-Fi mesh-networks, and governance issues, such as engaging with policy makers regarding suitable spectrum allocations.

IEEE 802.11ax, known now as 'Wi-Fi 6', is the first amendment in the Wi-Fi family to go beyond small indoor environments, and aim to optimise its performance in large outdoor deployments. Although, it enhances the nominal data rate by 37% compared to Wi-Fi 5, it aims at providing 4x improvement in terms of throughput and spectrum efficiency in dense deployments, through new features such as Orthogonal Frequency Division Multiple Access (OFDMA), Multi-User MIMO (MU-MIMO), and spatial reuse. At the same time, Wi-Fi 6 is reducing the power consumption per device. Whereas 2.4 and 5 GHz frequencies are used by legacy Wi-Fi technologies, the deployment of Wi-Fi 6E specifically relates to the use of the new 6 GHz spectrum band which has already been assigned in frontier markets (e.g. USA, Korea, UK etc.) and is expected to receive similar allocation elsewhere (e.g. Europe).

Table 1 provides comparative technical information on recent Wi-Fi generations.

---

[8] For more information about IEEE P802, see https://www.ieee802.org/, and especially, P802.11 (https://www.ieee802.org/11/) which is the working group responsible for WLAN standards such as the Wi-Fi standard. For more information on the Wi-Fi Alliance see https://www.wi-fi.org/.

*Table 1 Technical capabilities across legacy and current wireless standards*

| Features | Wi-Fi 4 (802.11n) | Wi-Fi 5 (802.11ac) | Wi-Fi 6/ Wi-Fi 6E (802.11ax) |
|---|---|---|---|
| Data rate | Up to 600 Mbps | Up to 7 Gbps | Up to 9.6 Gbps |
| Carrier Frequency | 2.4, 5 | 5 | 2.4, 5, 6 |
| Channel Bandwidth | 20, 40 | 20, 40, 80, 80+80, 160 | 20, 40, 80, 80+80, 160 |
| Frequency multiplexing | OFDM | OFDM | OFDM and OFDMA |
| OFDM symbol time (µs) | 3.2 | 3.2 | 12.8 |
| Guard interval (µs) | .04, .08 | .04, .08 | .08, 1.6, or 3.2 |
| Total symbol time (µs) | 3.6, 4.0 | 3.6, 4.0 | 13.6, 14.4, 16.0 |
| Modulation | BPSK, QPSK, 16-QAM, 64-QAM | BPSK, QPSK, 16-QAM, 64-QAM, 256-QAM | BPSK, QPSK, 16-QAM, 64-QAM, 256-QAM, 1024-QAM |
| MU-MIMO | N/A | DL | DL and UL |
| OFDMA | N/A | N/A | DL and UL |
| Radios | MIMO (4x4) | MU-MIMO (DL) (8x8) | MU-MIMO (DL & UL) (8x8) |

Four example scenarios for Wi-Fi 6 deployments include (Merlin, 2015):

1. Residential, where the deployment of Access Points (APs) is uncontrolled and unmanaged, resulting in high interference between the APs.
2. Enterprise, with low interference between the APs as the deployment is now managed and controlled.
3. Indoor small hexagon-based, representing the indoor dense scenarios (i.e. stadiums, auditorium etc.) where there is strong interference between the APs.
4. Outdoor large hexagon-based to assess the performance in outdoor hotspot deployments.

The deployment of Wi-Fi APs in these scenarios has traditionally been based on a relatively simple 'plug and play' setup for a single piece of equipment. Increasingly the use of Wi-Fi mesh systems has become popular, whereby rather than a single AP, the system consists of a main hub with multiple linked nodes spatially distributed throughout a building or home which are capable of capturing and rebroadcasting information (Navío-Marco et al., 2019). Such an approach helps to eliminate areas with poor signal coverage, improving both speed and reliability for users. The ability to easily deploy Wi-Fi strongly contrasts with the technical requirements of 5G deployment (Forge and Vu, 2020).

A variety of technical features have been introduced into the design of Wi-Fi 6 to help cope with the challenges of delivering consistent wireless connectivity, most of which focus on improving spectral efficiency and overall throughput, while still ensuring backward compatibility with previous generations. These include OFDMA, 1024 Quadrature Amplitude Modulation (QAM), uplink MU-MIMO, dual subcarrier modulation, spatial reuse ('BSS colouring'), and power saving technologies (to enhance energy efficiency).

One of the most important changes is the adoption by Wi-Fi 6 of mandatory support of OFDMA in both the downlink (DL) and uplink (UL). In contrast to cellular communications, where OFDMA is already in use, this is the first time this technical feature has been introduced to the Wi-Fi family. Indeed, OFDMA is one of the two techniques that allow Multi-User (MU) transmissions, where an AP can simultaneously transmit and receive information, to and from multiple users, in the same DL/UL. Like the legacy approach of OFDM, where the entire bandwidth is divided into multiple subcarriers, OFDMA allocates groups of these subcarriers, known as Resource Units (RUs), to different users, each one of them using different Modulation and Coding Scheme (MCS) and/or Transmit Power. Hence, OFDMA can improve spectral efficiency and provide up to 4x throughput gain when compared to OFDM by allocating either single or multiple RUs to users based on their needs (e.g. data to transmit) and the available channel conditions (e.g. SINR) (Khorov et al., 2019).

The second key MU technique is the support of MU-MIMO in both the DL and UL. Essentially, MIMO enables data transfer to take place across multiple antennas to take advantage of 'multipath propagation,' which is a technique for increasing the rate of transmission using different spatial streams of data. Although previous versions of Wi-Fi contain MIMO, Wi-Fi 6 enables multiple simultaneous beams (up to 8) to be supported by an AP, connecting to several devices concurrently (for both DL and UL). Additionally, when the signal quality is sufficiently good (i.e., high SINR), Wi-Fi 6 provides support for ultra-high capacity modulation technology that enables more bits to be packed into each Hz of frequency, thereby contributing to spectral efficiency and more efficient data transfer capabilities.[9]

To further reduce power consumption, Wi-Fi 6 adopts the power-saving technique introduced in IEEE 802.11ah, namely Target Wake Time (TWT). In contrast to the power-saving mechanisms introduced in the previous Wi-Fi generations, TWT allows devices to increase their sleep time, instead waking up at a specified time slot previously agreed to exchange data with the AP or other users.

To cope with the challenges in dense deployments where multiple Basic Service Sets (BSSs) might be operating on the same channel, the Spatial Reuse (SR) mechanism is introduced in Wi-Fi 6. The core functionality of this new feature is the 'BSS Colour', a 6-bit value carried on the physical header that aims to assist the devices to early identify whether a frame is an inter-BSS or an intra-BSS. Hence devices can abandon the reception of an inter-BSS frame, based on the interference level, to initiate a transmission to their BSS. This can increase the number of concurrent transmissions in a network, providing a throughput gain of 30% in outdoor dense deployments (Selinis et al., 2016).

---

[9] That is, Wi-Fi 6 supports ultra-high capacity 1024-QAM, compared to 5G, which only supports a maximum of 256-QAM. For more information, see https://www.commscope.com/blog/2018/wi-fi-6-fundamentals-what-is-1024-qam/.

Future generations of Wi-Fi are anticipated to support newer usage cases with ever stricter Quality of Service (QoS) requirements in terms of latency and throughput such as to support 8k video, holographics, etcetera. IEEE 802.11be (Lopez-Perez et al., 2019), which will be the successor of Wi-Fi 6, is expected to enhance throughput by at least 3-4 times, while maintaining backward compatibility with Wi-Fi 6. Support for larger channels (from 160 MHz to 320 MHz) and the increase in the number of spatial streams to 16 will boost the peak data rates to 30 Gbps. Multi-band aggregation, where channels in different frequency bands could be aggregated and used for data transmissions, is also under consideration for Wi-Fi 7.

Currently Wi-Fi 6 and earlier generations of Wi-Fi are half-duplex systems due to the challenges that full-duplex communications pose and to keep manufacturing costs low. However, Wi-Fi 7 aims at addressing these challenges by considering full-duplex systems. The coordination among the APs has also been proposed for Wi-Fi 7, to further utilise available resources and improve spatial reuse in dense deployments. Following cellular systems, the separation of data and control frames has been proposed, whilst the use of HARQ is also under consideration to provide reliable and low latency transmissions. Finally, Time-Sensitive Networking (TSN) technology could also be incorporated in Wi-Fi 7 to assure time synchronisation and low jitter (Adame et al., 2019), which has a significant impact on live-streaming applications.

Having now described the technical characteristics of both 5G and Wi-Fi, we will review the extent to which these technologies are similar or different with reference to wireless broadband connectivity.

## 5. Comparing and contrasting 5G and Wi-Fi 6

In this discussion, we compare 5G and Wi-Fi 6 on the basis of their technical characteristics, use of spectrum, business model and cost, and ease of installation and required skill level. These categories for comparison have been selected to account for both their engineering and economic relevance in the adoption of new technologies. As we have already discussed, both 5G and Wi-Fi 6 offer significant enhancements in performance with much faster connection speeds, higher device densities, and lower latencies, relative to prior technical generations.

The 5G and Wi-Fi 6 enhancements have narrowed the disparities in the legacy use cases that the cellular and Wi-Fi technology families may each address, while at the same time created opportunities to address new market demands for applications with ever-more intensive throughput and QoS requirements. The first effect tends to bring the two technologies closer as potential substitute alternatives for meeting end-user demand.[10] The latter effect may accentuate the importance of the fundamental differences that continue to differentiate cellular and Wi-Fi based networks, which would tend to move the two technologies apart as substitutes for demand.[11] At the same time, changes in the marketplace and enhancements in the capabilities upstream and downstream of the RAN (in the end-user devices, backbone networks, and from higher-layer protocols) make it easier to mix-and-match the technologies or compensate for perceived deficits in either technology, based on the needs of a particular context. From the perspective of network deployers (whether those be MNOs, new types of service providers, or end-users), 5G and Wi-

---

[10] For example, 5G enhancements allow 5G to better address standalone or indoor deployments – traditional markets historically best served by Wi-Fi technologies. At the same time, Wi-Fi 6 enhancements coupled to the global expansion in the availability of Wi-Fi APs (supporting earlier generations of Wi-Fi) are enabling Wi-Fi to be seen as a viable competitor for wider-area coverage wireless connectivity needs. In this way, 5G and Wi-Fi 6 enhancements render the cellular and Wi-Fi technologies closer substitutes for meeting the needs of end-users and providers of network services.

[11] In the case of cellular and its latest incarnation, 5G, that includes better support for managing multiple APs and the needs of per-connection QoS management support, but at the cost of additional RAN overhead. In the case of Wi-Fi and its latest incarnation, Wi-Fi 6, that includes faster raw throughput with significantly less RAN overhead.

Fi 6 enhancements render the cellular and Wi-Fi families of technologies as "supply" tools in an expanded tool-box to address end-user "demand" that is increasingly context-dependent. A simple characterisation of either as a substitute for the other or as a complement is inappropriate, since they may play either role in different situations.

In the following, we disentangle some of these conflicting forces to better understand how 5G and Wi-Fi 6 may alter the competitive dynamics between cellular and Wi-Fi-based technologies. Table 2 provides a high-level summary comparison of 5G and Wi-Fi 6 for multiple engineering and economic dimensions.

*Table 2 Comparing key 3GPP 5G and Wi-Fi 6 (IEEE 802.11ax) features*

| Category | Variable | 3GPP 5G | Wi-Fi 6 / Wi-Fi 6E |
| --- | --- | --- | --- |
| Technical | Peak data rate | 2 Gbps (DL), 1Gbps (UL) | 10 Gbps 8x8 (DL), 5 Gbps (UL) |
| Technical | MU-MIMO | 128x128 | 8x8 |
| Technical | Coverage range | 100-300 meters for small cells, up to tens of km for macro cells | <50 meters indoor, up to 300 meters outdoor |
| Technical | Carrier aggregation | Yes | Yes, 40, 80, 160 (or 80+80) |
| Technical | Inter-cell interference | Controlled | Mainly uncontrolled |
| Technical | Channel Access Scheme | OFDMA | OFDMA |
| Spectrum | License type | Mostly licensed | Unlicensed |
| Spectrum | General bands | Low, mid and high | Low and mid |
| Spectrum | Specific frequencies | Low-band (<1 GHz), mid-band (1-7 GHz) and high-band (~24-29 GHz) | 2.4 GHz, 5 GHz, 6 GHz, 60 GHz |
| Spectrum | Channel Bandwidth | 20, 40, 80, 100 MHz | 20, 40, 80, 160 MHz |
| Business model and cost | Revenue model | Pre- or post-pay billing for data services | Either a service, 'free', amenity, or pure WLAN without external connection |
| Business model and cost | User equipment price | High | Low |
| Business model and cost | Public versus private | Traditionally publicly provided by an MNO | Traditionally privately provided |

| Business model and cost | Chip/modem cost | High | Low |
|---|---|---|---|
| Business model and cost | Data cost | Monthly subscription ($5-20) | Free ('piggybacks' on fixed broadband) |
| Installation and skills | Deployment approach | Controlled and managed | Uncontrolled and mostly un-managed |
| Installation and skills | Installation skill level | High | Low |
| Installation and skills | Development skill level | High | Low |

Many of the features highlighted in Table 2 are not unique to the latest incarnations of the cellular and Wi-Fi technology families, but are the result of decades of evolution. Consideration of this background assists in illuminating the implications of these differences. For example, legacy cellular networks were designed to provide wide-area connectivity for large numbers of users roaming across vast coverage areas. This was most efficiently supported with higher-power, macro cell architectures that could provide single-cell coverage. This approach could reduce the need for high-speed cell hand-offs and reduce wide-area costs. However, the drawback of a macro cell design is the limited per-user capacity when compared to the peak capacity available via Wi-Fi 6 or smaller cell 5G deployments. A new approach to deal with this issue in wide-area networks (e.g. dense urban or rural scenarios) is to use much higher order MU-MIMO (128x128) when compared to Wi-Fi 6 (8x8). Another is to for MNOs to shift towards ever-smaller cell architectures to gain the capacity benefits of network densification. This flexibility assists MNOs in integrating 5G with MNO carriers' macro cell networks, and helps them rapidly provide wider-area coverage with scalable capacity as small cells are built out (first in high-demand locations, and potentially later as supporting infrastructure such as backhaul connectivity is built out). As higher power is allowed in licensed bands, current 5G small cells target larger coverage areas (100-300 meters) than Wi-Fi cells (e.g. <50 meters indoors). Both technologies take advantage of carrier aggregation and OFDMA as the main channel access scheme to provide greater capacity to users via increased spectrum agility (Chavarria-Reyes et al., 2016).

Traditionally, Wi-Fi technologies have operated at reduced power, offering limited range coverage for each Wi-Fi AP. This small cell architecture is consistent with the design goals of Wi-Fi as a WLAN technology operating using unlicensed spectrum, benefiting from frequent spectral reuse to provide very high capacity local Internet connectivity (Up to 10 Gbps in Wi-Fi 6). By definition, a Local Area Network (LAN) is intended to provide coverage for a relatively small contiguous geographic area, but via gateways and repeaters, the range of WLANs can be extended to ever-larger coverage areas such as a corporate or academic campus locations. Additionally, consistent with the unmanaged and uncontrolled end-user-based deployment model for WLANs, the goal in Wi-Fi 6 is to continue to provide local wireless connectivity to a relatively small community of users and devices over a contiguous area, most typically indoors but connectivity is readily expanded outdoors in campus environments. Small cells, by their nature, support a smaller number of users and when deployed indoors (or on a campus), make it easier to manage shared connectivity to avoid destructive interference among users and to avoid interference from WLANs deployed by other network operators. Taken together, these WLAN usage requirements avoid the need for including extensive capabilities to manage large numbers of APs and enabling high-speed hand-offs as users move across coverage areas of adjacent APs. This contrasts with the approach in 5G where inter-AP interference mitigation and coordinated operation is intrinsic to the design of the technology and the service architecture.

After the first AP (whether Wi-Fi or cellular), the interconnection to wide-area networks or other APs may be via wired (often fibre) backhaul connections, or if small cells are deployed densely enough, via wireless backhaul. This means that both the network backhaul connection and last-hop AP performance of Wi-Fi 6 and 5G networks may be quite similar so long as the usage-case does not call for supporting fast-movement (e.g. at highway or airplane speeds), necessitating rapid hand-offs to adjacent APs. Supporting such applications was a focal requirement for the design of cellular technologies, including 5G. In contrast, the mobility support for Wi-Fi was based on supporting *nomadic* use, where users move between high-capacity hotspots, but generally do not expect to remain seamlessly connected in transit between hotspots (although this can happen at slow speeds on campus networks). The question is how much of future usage will fit with the nomadic mobility model, with many users being quasi-fixed, as opposed to requiring fast-mobile connectivity. As highlighted earlier in this paper, most existing data traffic demand is produced in-homes, within range of in-home Wi-Fi hotspots using unlicensed spectrum. Moreover, although Wi-Fi deployments often consist of one or only a few APs, in private networking contexts where the deployment of unaffiliated WLANs may be controlled, contiguous Wi-Fi coverage and support for low-speed roaming among APs is readily implemented for entire building complexes and campus environments.

With the new developments in these technologies there is likely to be a shift in the loci of control. Often this is characterised as 'convergence', where the two technologies are blended to take advantage of the unique and complementary capabilities of both access networks to provide 'seamless' network services. Or 'divergence' where connectivity via each technology moves towards monolithic connectivity domains. For example, an ongoing trend is the significant proportion of cellular traffic that is opportunistically off-loaded to Wi-Fi hotspots that an end-user's wireless device may have access to (either under service provider or end-user control of the off-load behaviour). There are multiple options for controlling how the off-loading is managed, including how to authenticate, route traffic (e.g. through the operator's mobile core or straight to the 'Internet'), and manage connection control (e.g. roaming). Traditionally cellular operators attempting to utilise Wi-Fi to offload traffic must consider the QoS implications which could lead to a poorer user experience if delivered incorrectly. The potential convergence of 5G and Wi-Fi 6 is viewed by some to enable new use cases and thus future business opportunities including for enterprise networking, 'factories of the future', smart cities, and home connectivity (WBA and NGMN Alliance, 2019). The enhancements to Wi-Fi 6 better enable Wi-Fi networks to provide the higher, enterprise-grade Quality of Experience required both when deployed by mobile operators (legacy or new) or by users themselves.

The context-dependence of how to best mix-and-match cellular and Wi-Fi is affected by where the network decision-maker sits with respect to its current network capabilities (e.g., service provider, with cellular, Wi-Fi or mixed RAN; end-user seeking to self-provision or outsource) and the demand scenario to be addressed (e.g., wide-area or local, QoS requirements, etc.). These differences feed into the choice of where to locate control and the most suitable technology option to select. For example, there is debate around whether Wi-Fi should be 'anchored' in a 3GPP control layer, or if it should be independently controlled by user, application, operating system or $3^{rd}$-party connection manager. Generally, Wi-Fi control resides with the owner of the device, whereas cellular control resides with the owner of the SIM. In recent years there have been new developments such as Google's experimental Project Fi (now Google Fi) which acts as a Mobile Virtual Network Operator (MVNO), aiming to dynamically switch users to the best connection available (based on signal and speed) whether from cellular or Wi-Fi. While Wi-Fi may be the preference, if a user loses Wi-Fi connectivity Google Fi automatically switches the call to cellular. Although this is an innovative concept, the current user base makes this more of a niche service with current indications suggesting that this is unlikely to change. Acting as an MVNO which preferences Wi-Fi is a starting point to become a more robust competitor to traditional MNO business model approaches.

While fast mobility will remain important, we anticipate that much of the growth in traffic and usage models for both 5G and Wi-Fi 6 will be associated with quasi-fixed, nomadic usage cases. The 5G standards are anticipating this and future standards will enable standalone 5G APs and unlicensed spectrum connectivity to compete directly with Wi-Fi WLANs. Alternatively, while Wi-Fi WLANs were originally targeted principally at single AP WLANs or WLANs consisting of a relatively small number of APs in a local, geographically contiguous area to support hotspot coverage, the 802.11 protocols have been expanded to support management of greater numbers of APs over a larger area and to support higher-speed AP hand-offs. For example, Dedicated Short-Range Communications using IEEE 802.11p was developed for use in automotive applications (Katsaros and Dianati, 2017; Mir and Filali, 2014). Hence, vehicular networking is another battleground for the two technologies, with the 802.11p competing with an LTE 4G/5G cellular vehicular networking standard called Cellular V2X (Mir and Filali, 2014).

The type of spectrum used in legacy architectures also has had an important impact on the design of wireless technologies and has helped shape the evolution of cellular and Wi-Fi technologies. For example, we have already highlighted that cellular has a history of providing wide-area support for fast-moving mobile users, with longer-range macro cells mounted on high towers, using licensed spectrum (mostly acquired via auction). Large up-front and continuing capital investments are needed to deploy wide-area coverage networks and acquire the requisite spectrum licenses before service revenues are obtained, and those investments need to be amortised over many years. Licensed frequencies have enabled MNOs to have predictable spectrum quality since they can manage how users (i.e., MNO subscribers) share available spectrum resources. Hence, initially 4G, but 5G over the long term, is likely to be the preferred choice for supporting applications which fit this niche, such as providing connectivity to autonomous vehicles, drones, and enabling those smart cities or IoT applications which require ubiquitous connectivity over wide areas (Oughton and Russell, 2020). In contrast, Wi-Fi has traditionally used unlicensed spectrum which is free to use but is shared non-cooperatively with other local users competing for spectrum.

Additionally, legacy MNO networks focused on supporting mobile telephony relied on paired bands using FDD to allow symmetric uplink/downlink channels, whereas WLANs relied on a single shared spectrum band that was better suited to the asynchronous, variable rate traffic typical of data networks. However, in cellular this is now starting to change. For example, as cellular traffic becomes more variable rate and heterogeneous, the trend is toward increasing usage of unpaired spectrum bands which can utilise TDD. This is possible in the 3.5 GHz band which is central to delivering high capacity 5G but will also be widely used at millimetre wave frequencies (e.g. the 26-28 GHz band) (Oughton et al., 2017).

Significant new licensed and unlicensed spectrum is now being made available for both 5G and Wi-Fi 6 deployments. For licensed spectrum, new allocations of high-band frequencies above 28 GHz will be auctioned along with prime, mid-band spectrum in the 3-5 GHz range (often used by satellite broadcasters). For unlicensed spectrum, in the U.S., the Federal Communications Commission has allocated 1200 MHz for unlicensed use in the 6 GHz band, and other countries are following but with debate around the size of the allocated bandwidth. It is anticipated that both the next generation of 5G that will enable standalone 5G deployments (i.e., use of unlicensed spectrum without requiring control via a licensed band) and Wi-Fi 6 will compete head-to-head to co-exist and share the 6 GHz unlicensed spectrum.

Traditionally over the past two decades there have been quite clear demarcations between public and private networks. However, changing norms and regulations for spectrum usage are beginning to blur these boundaries (Disruptive Analysis, 2020). For example, spectrum policymakers are looking to expand management options to enable greater sharing among heterogeneous spectrum users, including government and commercial users (Massaro, 2017; Massaro and Beltrán, 2020; Saint and Brown, 2019; Sohul et al., 2015). A noteworthy example is the model for sharing 3.5 GHz spectrum in the newly enabled Citizens Band Radio Service (CBRS), that began operations in the U.S. in the fall of 2019. The CBRS allows multiple tiers of priorities of users (i.e., legacy government users, new priority commercial and unlicensed commercial users) to share the spectrum according to dynamic control of tiered interference protection rights (Grissa et al., 2019; Souryal and Nguyen, 2019; Yrjölä and Jette, 2019). Equally, other countries have displayed leadership in providing localised spectrum licensing including the UK, Japan, Germany and France. Additionally, to adjust to the higher capital costs associated with smaller cells (which are also needed to make use of high frequency millimetre wave spectrum), regulators and MNOs are looking towards new shared spectrum usage models (Gomez et al., 2019, 2020; Weiss et al., 2019; Weiss and Jondral, 2004). Such changes provide greater efficiency, by allowing users to access a greater quantity of the spectrum resources available. Figure 3 illustrates how the types of wireless services which can be deployed are changing, partially driven by the availability of local and/or shared spectrum resources.

*Figure 3 Shifting public-private boundaries towards hybrid networks*

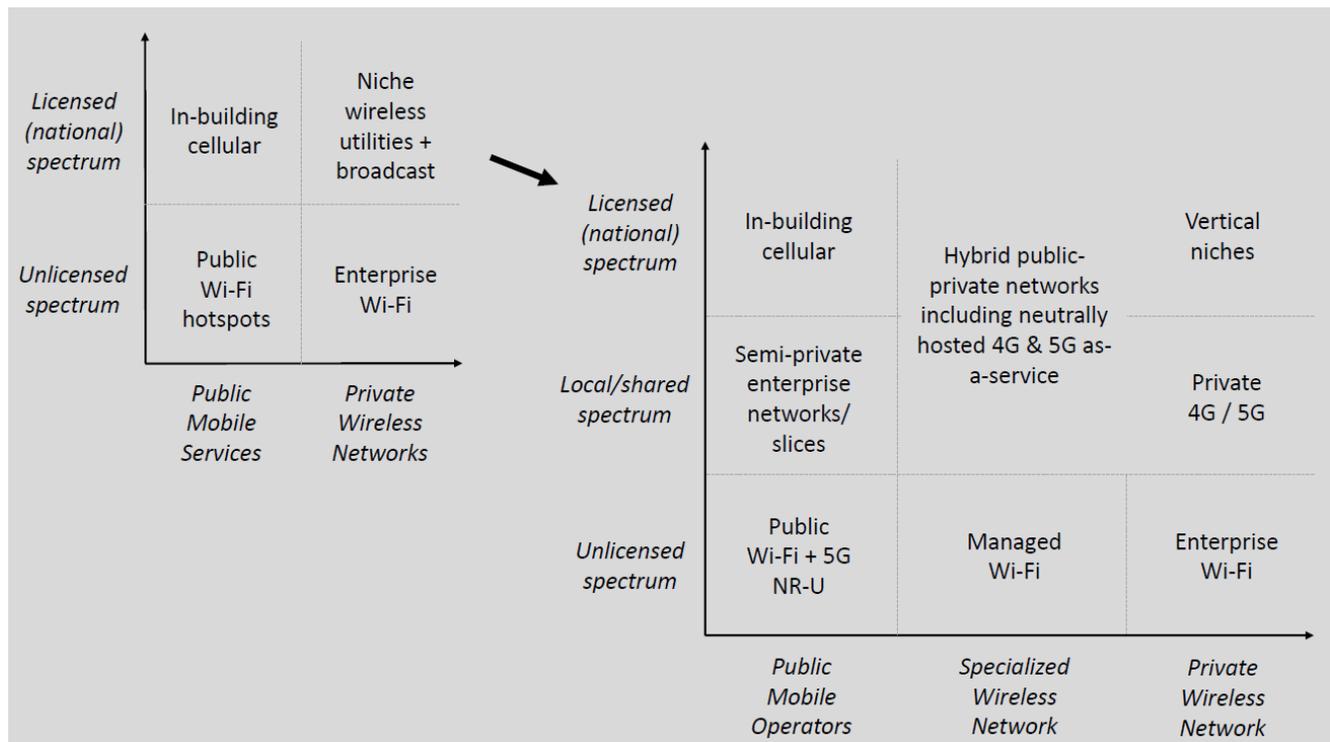

Such spectrum policy changes have led to a variety of new hybrid network deployment models. For example, 'semi-public' or 'semi-private' networks are starting to emerge, such as 4G/5G networks run by enterprises or specialised Business-to-Business or wholesale MNOs. Whereas in previous decades there were well defined boundaries between public and private networks, using either unlicensed or licensed spectrum, this strict delineation is fading. Whereas cellular was traditionally based on licensed bands, and Wi-Fi based on using unlicensed bands, there is now a hybridisation in the provision of private cellular networks which can take advantage of either unlicensed or local and shared spectrum, provided by specialist communications providers (so not by MNOs or the enterprise which takes advantage of the provided services). Despite this supply-side shift, many technologies are likely to co-exist with users having multiple devices and each device having multiple radios. Thus, depending on the availability of wireless services, devices may simultaneously or dynamically make use of both private or public cellular (4G/5G), and Wi-Fi networks (Wi-Fi 5/6) both in and outside of homes or businesses.

In terms of business model and cost, Wi-Fi 6 may have an advantage for indoor and private local network deployments. This arises because of its legacy as the technology choice for WLANs due to the low cost and scalable deployment of IEEE 802.11. Historically, end-users could deploy WLANs with a few APs using off-the-shelf, inexpensive Wi-Fi equipment that operates in unlicensed spectrum, leading to a lower per square meter cost than cellular (Intel, 2020; TechRadar, 2019). These WLANs provided local wireless connectivity to shared fixed access broadband in the home, office, or coffee shop. Wi-Fi 6 offers an enhanced WLAN and so may be the preferred technology of choice for connecting all IoT devices around the home, from laptops to security cameras and home appliances.

As the supporting networks have become more flexible, and the ability of the cellular and Wi-Fi technologies have become more capable with 5G and Wi-Fi 6, underlying differences in the prices (costs) of the chipsets for the technologies have become increasingly important. While publicly available data on the prices for cellular chips is scarce (due to commercial sensitivities), the approximate costs are reported by media outlets covering the tech industry. For example, a Qualcomm basic 5G chipset (Snapdragon

765) ranges from $25-40, with the top-of-the-line Snapdragon 5G modem costing $120-$130 (Friedman, 2020). Moreover, an entry level 5G MediaTek chip is approximately $40, increasing up to $60-$70 per chip for a flagship 5G product (Dimensity 1000). In comparison, a Wi-Fi 6 chip is significantly cheaper, with purchase prices in the range of $12-18 as of the end of 2020, for example, for a Qualcomm Wi-Fi 6 chip (Qca6391) (Alibaba, 2020a, 2020b). Many Wi-Fi consumer IoT devices aim for a price range of $50-200 (Amazon, 2020), therefore adding a 5G chip/modem is a significant cost and could affect product viability. In contrast, 5G smartphone device costs range from approximately $300-1200. With Wi-Fi 6 based on IEEE standards the per-device cost for the associated licenses is dramatically lower than for cellular products, enabling manufacturers to build an entire computer with a built-in Wi-Fi radio for under $20, whereas simply integrating a cellular modem adds more than $100 to the price of a device (Cisco, 2019).

Furthermore, the relative costs of using cellular (5G or earlier) or Wi-Fi (Wi-Fi 6 or earlier) is highly path dependent for individual network adopters.[12] For example, with most existing smart home devices such as TVs and voice assistants using Wi-Fi, with practically none using cellular, making a full shift to cellular look extremely unlikely.

---

[12] Path dependence has been an active area of research for over 30 years (Arthur, 1994; David, 1985) and occurs when historical events continue to have serious ramifications for future decisions. For example, often one technology can become the dominant standard over another, not because of better technical specifications, but due to *serendipity*. However, once momentum behind a technology is gained, it can be very difficult to switch paths to another, often due to the economics of increasing returns to scale. Such examples include the design of the QWERTY keyboard, or the videotape format war between VHS and Betamax. In the wireless domain analogous effects have occurred with the historical inclusion of Wi-Fi in laptops, with both Apple & Intel being drivers of this dating back to the era of the Centrino chip. This is the path-initiator which led consumers to expect to have access to Wi-Fi in computing devices and for the feature to be controlled directly by themselves, without the need for a service provider.

Importantly, the economic outlook for the mobile telecommunication sector is poor with ARPU either static or declining in many countries, which may make it hard to deliver on the high societal expectations of new technologies such as 5G. With falling data prices and exponentially increasing traffic growth, MNOs are anxious to tap into new sources of revenue. MNOs are hoping 5G will allow them develop new revenue opportunities in industrial IoT and other vertical sectors such as energy, health and automotive. This is part of the specialist communication services mentioned previously. While most businesses already use Wi-Fi, some may now have the option to choose which wireless connectivity technology best suits their needs. For example, Wi-Fi 6 would be the natural successor to most existing networks providing a low cost, scalable option for uses with low Quality of Service requirements. However, for automation at a factory or campus with very high Quality of Service requirements (e.g. latency, 99.9999% reliability etc.), either working with an MNO or gaining locally licensed spectrum for private 4G LTE or 5G bands might offer a better option by ensuring dedicated access to spectrum resources (Matinmikko et al., 2018). Ultimately these decisions will be very application and sector specific and will also depend on the level of mobility, area needing wireless coverage and cost. For static uses (e.g. machinery) a fixed fibre connection could be the best option, but a factory using moving robots may be more suited to a private network. For example, the UK food distributor Ocado currently uses unlicensed cellular networks in their factory automated robotics but switching to privately licensed spectrum could be a future development. Given these circumstances, it is hard to see how either cellular or Wi-Fi will dominate over the other given the range of different requirements each use case has. Indeed, both wireless connectivity technologies may also face competition from fixed connectivity if no mobility is required.

## 6. Discussion and conclusions

Herein we revisited the debate associated with wireless Internet connectivity by providing a new evaluation of the two main technologies involved in the provision of next generation wireless broadband: 5G and Wi-Fi 6. Our analysis highlights how the futures for 5G and Wi-Fi 6 needs to be understood within the larger context of how earlier generations of cellular and Wi-Fi technologies have shaped the evolution of wireless networking and what this may mean for the future.

First, in terms of general demand-side trends, data traffic is expected to continue to grow significantly with an increasing proportion of devices utilising wireless connectivity as the first connection point. The COVID-19 pandemic of 2019-2021 has highlighted the importance of enhanced digital connectivity to support remote work, education, and social engagement during the global crisis. But there may also be potentially new trends which could arise out of the shifting work and social patterns produced by the pandemic. Such changes could have repercussions for the spatial and temporal usage of wireless broadband connectivity and the associated economics of each technology. Additionally, the ongoing consumption of ever-more and ever-higher quality video content will also be an important factor driving consumer data demand, while enterprise use will reflect growing adoption of cloud-based applications and computing platforms.

A key goal of this analysis was to highlight how 5G and Wi-Fi 6 may affect the competitive dynamics between the cellular and Wi-Fi technology families. To determine whether it looks as if one may emerge as the dominant technology choice or whether both technologies would continue to be important and often complementary tools for meeting wireless needs in the future. The previous competition between 3G and Wi-Fi was shaped by the networking challenges of an earlier era, but now this landscape has shifted considerably. The evaluation undertaken here demonstrates that while each of the technologies has relative advantages stemming in large measure from their different legacy trajectories and focal usage scenarios, the two technologies will find themselves appropriately viewed both as alternative and substitute options for many contexts, as well as complements in many others. In this section we discuss and conclude these findings.

In terms technical characteristics, both new generations of cellular and Wi-Fi aim to provide more spectrally efficient radio interfaces to support a better user experience. But we find that generally 5G is still focusing on delivering high mobility to users, as with previous cellular generations. While Wi-Fi remains aimed at providing nomadic high-capacity hotspots which can be easily deployed. Meanwhile, 5G is allowing the next generation of cellular technology to target new private and standalone networking opportunities, especially in industrial vertical sectors, that were previously the niche of a wide variety of legacy Wi-Fi or other proprietary radio systems. For example, while the main difference is generally the use of licensed rather than unlicensed spectrum, there is now even a standard which enables 5G to operate in unlicensed bands (5G NR-U). At the same time, the growth of quasi-nomadic usage and the expansion of small cell deployments is allowing wider-area network providers (like wired broadband providers) to expand into mass-market mobile services. 5G's new 'network slicing' mechanisms, together with extra specifications for verticals, also should enable customised virtual networks to have specific capabilities for enterprise, especially from 3GPP Release 17 onwards.

Changes in spectrum policy have had a substantial impact. For example, the introduction of private spectrum licensing regimes for local areas has opened new opportunities for specialist communications providers to deploy cellular networks in private enterprises. Therefore, while much of the rest of the consumer telecommunications landscape is moving towards increased infrastructure convergence, with fixed operators selling mobile and vice versa, the opposite is true for business wireless connectivity. Rather than a shift towards more centralised monolithic communications providers, there is divergence driven by the need for many specialist providers to deliver bespoke private, semi-private and neutrally hosted 4G and 5G networks for different industrial sectors.

These changes mean that the use of industrial IoT across a range of manufacturing and warehouse facilities leaves firms with new options. They can choose to outsource their networking needs to an MNO or newer specialised provider, and those may offer a variety of options. Moreover, if they elect self-provision (because of the control or perceived cost benefits of such a choice), industrial users will have additional options. Self-provision could take place by either continuing to use existing Wi-Fi connectivity, deploy a private enterprise 4G or 5G network with locally licensed spectrum, or take advantage of both technologies. Indeed, the range of options has expanded and become more scalable. If industrial users select a private cellular network, that will provide a high degree of control over the provision of wireless connectivity with strict Quality of Service requirements (e.g. 99.9999% reliability etc.), thanks to exclusive access to locally licensed spectrum resources. Such an opportunity could be highly useful for automating processes which require reliable wireless connectivity (e.g. for mobile robots). The benefit of remaining with Wi-Fi is the ability to cheaply and quickly provide wireless Internet connectivity to traffic which does not require high Quality of Service. In the past, over-provisioning for capacity often proved adequate to ensure the requisite level of service quality, and in closed spectrum environments (indoors, campus environments etc.) the risk of interference from unaffiliated radio networks may be minimal. We expect businesses that decide to deploy their own private cellular network are highly likely to continue using Wi-Fi simultaneously, suggesting that these technologies will remain complements to each other for the foreseeable future.

For non-industrial sectors such as enterprise offices and retail sites, or for visitor-heavy venues such as hotels and airport terminals, there will be parallel needs for Wi-Fi and public cellular access, although private cellular may have less impetus for deployment.

Will 5G 'kill-off' Wi-Fi? This is one of the main questions which motivated this analysis, given the ongoing debate in industry on this topic. Ultimately, the competition between 5G and Wi-Fi 6 technologies offers important benefits by enabling greater flexibility for users to mix-and-match the technologies, business models, and spectrum usage models to best fit their needs. Proponents of one or the other technology, however, may argue for the benefits of their chosen technology displacing the other, and may argue for regulatory policies that would serve to tilt the marketplace in their favour. We believe such efforts need to be resisted, and that both technologies have important roles to play in the marketplace, based on the needs of different use cases. This is particularly important given that apart from smartphones, some devices will remain Wi-Fi-only, while some cellular-only, with just a fraction actively using both technologies to steer traffic based on user preference. Additionally, we expect cost economics and convenience of deployment to play a major role.

Given the path dependence exhibited by sunk costs in legacy infrastructure, it is unlikely that either technology will be able to usurp the other due to the additional costs of transitioning, except in a few specific circumstances. For example, cellular will remain the dominant wide-area technology thanks to the sunk investments made in existing brownfield infrastructure (towers, backhaul fibre etc.) which can be reused to provide generational upgrades at a lower cost than new greenfield deployments. Equally, it is hard to see how Wi-Fi would be threatened by 5G cellular for indoor locations, particularly for homes, given the ongoing challenges cellular technologies have with trying to serve inbuilding users with a high degree of reliability. If wireless devices do not require mobility or high Quality of Service, it is hard to find a justification for using 5G given it is generally more expensive, particularly for consumer electronics. Certainly, cost economics will be a major factor which affects the design of wireless devices, as well as consumer behaviour, and Wi-Fi has the advantage in this area, even if cellular is moving towards unlimited data subscriptions.

Future research should consider evaluating the historical rivalries between these two technology types, cellular and Wi-Fi, in order to understand how past decisions have placed wireless broadband connectivity options on a path dependent trajectory.

The contribution of this paper has been to consider the future evolution of the wireless broadband landscape over the next decade, as 5G and Wi-Fi 6 begin to roll-out and are adopted. Surprisingly, such discussion has been widespread in industry, but not yet comprehensively evaluated in academic terms, justifying the contribution of this manuscript to the telecommunications policy community. With the ongoing blurring of boundaries between 5G cellular and Wi-Fi 6, such evaluation will need to continue to understand how the competitive dynamics of these technologies play out for each consumer and industrial use case.

# 7. References

3GPP, 2020. 3GPP Release Timeline [WWW Document]. URL https://www.3gpp.org/ (accessed 12.24.20).

3GPP, 2016a. 3GPP TR 22.891 V14.2.0 (2016-09). Technical Specification Group Services and System Aspects; Feasibility Study on New Services and Markets Technology Enablers; Stage 1 (Release 14). 3rd Generation Partnership Project, Valbonne, France.

3GPP, 2016b. 3GPP TR 22.863. Feasibility study on new services and markets. Technology enablers for enhanced mobile broadband. Stage 1 (Release 14). 3rd Generation Partnership Project, Valbonne, France.

3GPP, 2016c. 3GPP TR 22.862. Feasibility study on new services and markets. Technology enablers for critical communications. Stage 1 (Release 14). 3rd Generation Partnership Project, Valbonne, France.

3GPP, 2016d. 3GPP TR 22.861. Feasibility study on new services and markets. Technology enablers for massive internet of things. Stage 1 (Release 14). 3rd Generation Partnership Project, Valbonne, France.

3GPP, 2016e. Specification # 38.801. Study on new radio access technology: Radio access architecture and interfaces. Version 14 (No. Version 14). 3rd Generation Partnership Project, Valbonne, France.

5G PPP Architecture Working Group, 2019. View on 5G Architecture (No. Version 3.0). European Commisson, Brussels, Belgium.

Adame, T., Carrascosa, M., Bellalta, B., 2019. Time-Sensitive Networking in IEEE 802.11 be: On the Way to Low-latency WiFi 7. arXiv preprint arXiv:1912.06086.

Akdeniz, M.R., Liu, Y., Samimi, M.K., Sun, S., Rangan, S., Rappaport, T.S., Erkip, E., 2014. Millimeter wave channel modeling and cellular capacity evaluation. IEEE journal on selected areas in communications 32, 1164–1179.

Alibaba, 2020a. 5.8g Qualcomm Wifi 6 Chip Qca6391 M.2 802.11ax Wifi+bt 5.1 Module - Buy 5ghz Wifi Module,Qualcomm Module,Wifi Bt Module Product on Alibaba.com [WWW Document]. URL https://www.alibaba.com/product-detail/5-8G-Qualcomm-WiFi-6-Chip_62398834266.html?spm=a2700.details.maylikeexp.5.52191278lsocNP (accessed 12.27.20).

Alibaba, 2020b. Smt Version Qualcomm Qca6391 Wifi 6 Chip 2.4/5ghz Wifi Module For Outdoor Wifi Ap [WWW Document]. URL https://www.alibaba.com/product-detail/SMT-Version-Qualcomm-QCA6391-WiFi-6_1600100926830.html?spm=a2700.details.deiletai6.9.778a6385aDSz0X (accessed 12.27.20).

Alsharif, M.H., Nordin, R., 2017. Evolution towards fifth generation (5G) wireless networks: Current trends and challenges in the deployment of millimetre wave, massive MIMO, and small cells. Telecommun Syst 64, 617–637. https://doi.org/10.1007/s11235-016-0195-x

Amazon, 2020. Wi-Fi Enabled Devices [WWW Document]. Amazon Marketplace. URL https://www.amazon.co.uk/s?k=wifi+enabled+devices&ref=nb_sb_noss

Arthur, W.B., 1994. Increasing returns and path dependence in the economy. University of Michigan Press.

Bai, T., Alkhateeb, A., Heath, R.W., 2014. Coverage and capacity of millimeter-wave cellular networks. IEEE Communications Magazine 52, 70–77. https://doi.org/10.1109/MCOM.2014.6894455

Bauer, J.M., 2018. The Internet and income inequality: Socio-economic challenges in a hyperconnected society. Telecommunications Policy, SI: Interconnecting 42, 333–343. https://doi.org/10.1016/j.telpol.2017.05.009

Bloomberg, 2017. A World Without Wi-Fi Looks Possible as Unlimited Plans Rise. Bloomberg.com.

Boccardi, F., Heath, R.W., Lozano, A., Marzetta, T.L., Popovski, P., 2014. Five disruptive technology directions for 5G. IEEE Communications Magazine 52, 74–80. https://doi.org/10.1109/MCOM.2014.6736746

Bogale, T.E., Le, L.B., 2016. Massive MIMO and mmWave for 5G Wireless HetNet: Potential Benefits and Challenges. IEEE Vehicular Technology Magazine 11, 64–75. https://doi.org/10.1109/MVT.2015.2496240

Cave, M., 2018. How disruptive is 5G? Telecommunications Policy, The implications of 5G networks: Paving the way for mobile innovation? 42, 653–658. https://doi.org/10.1016/j.telpol.2018.05.005

Chavarria-Reyes, E., Akyildiz, I.F., Fadel, E., 2016. Energy-Efficient Multi-Stream Carrier Aggregation for Heterogeneous Networks in 5G Wireless Systems. IEEE Transactions on Wireless Communications 15, 7432–7443. https://doi.org/10.1109/TWC.2016.2602336

Cisco, 2020. Cisco Annual Internet Report - Cisco Annual Internet Report (2018–2023) White Paper [WWW Document]. Cisco. URL https://www.cisco.com/c/en/us/solutions/collateral/executive-perspectives/annual-internet-report/white-paper-c11-741490.html (accessed 7.27.20).

Cisco, 2019. Why Wi-Fi 6 and 5G Are Different: Physics, Economics, and Human Behavior - Cisco Blogs [WWW Document]. URL https://blogs.cisco.com/networking/why-wifi6-and-5g-are-different (accessed 12.27.20).

Claffy, K.C., Clark, D.D., Bauer, S., Dhamdhere, A., 2020. Policy Challenges in Mapping Internet Interdomain Congestion. Journal of Information Policy 10, 1–44. https://doi.org/10.5325/jinfopoli.10.2020.0001

David, P.A., 1985. Clio and the Economics of QWERTY. The American economic review 75, 332–337.

Disruptive Analysis, 2020. Private and semi-private wireless networks: A closer look at Private Networks in 2020 and beyond. A Disruptive Analysis thought-leadership eBook. Disruptive Analysis, London.

Eurostat, 2020. Digital economy and society statistics [WWW Document]. URL https://ec.europa.eu/eurostat/web/digital-economy-and-society/data/database (accessed 7.19.20).

Forge, S., Vu, K., 2020. Forming a 5G strategy for developing countries: A note for policy makers. Telecommunications Policy 44, 101975. https://doi.org/10.1016/j.telpol.2020.101975

Frias, Z., Mendo, L., Oughton, E.J., 2020. How Does Spectrum Affect Mobile Network Deployments? Empirical Analysis Using Crowdsourced Big Data. IEEE Access 8, 190812–190821. https://doi.org/10.1109/ACCESS.2020.3031963

Friedman, A., 2020. Top analyst says Qualcomm has started a price war that will impact prices of 5G phones [WWW Document]. Phone Arena. URL https://www.phonearena.com/news/qualcomms-price-war-could-harm-mediatek_id121536 (accessed 12.26.20).

Ge, X., Cheng, H., Guizani, M., Han, T., 2014. 5G wireless backhaul networks: challenges and research advances. IEEE Network 28, 6–11. https://doi.org/10.1109/MNET.2014.6963798

Ge, X., Tu, S., Mao, G., Wang, C.-X., Han, T., 2016. 5G Ultra-Dense Cellular Networks. IEEE Wireless Communications 23, 72–79. https://doi.org/10.1109/MWC.2016.7422408

Gomez, M., Weiss, M., Krishnamurthy, P., 2019. Improving Liquidity in Secondary Spectrum Markets: Virtualizing Spectrum for Fungibility. IEEE Transactions on Cognitive Communications and Networking 5, 252–266. https://doi.org/10.1109/TCCN.2019.2901787

Gomez, M.M., Chatterjee, S., Abdel-Rahman, M.J., MacKenzie, A.B., Weiss, M.B.H., DaSilva, L., 2020. Market-Driven Stochastic Resource Allocation Framework for Wireless Network Virtualization. IEEE Systems Journal 14, 489–499. https://doi.org/10.1109/JSYST.2019.2927443

Graham, M., Dutton, W.H., 2019. Society and the Internet: How Networks of Information and Communication are Changing Our Lives. Oxford University Press.

Grissa, M., Yavuz, A.A., Hamdaoui, B., 2019. TrustSAS: A Trustworthy Spectrum Access System for the 3.5 GHz CBRS Band, in: IEEE INFOCOM 2019 - IEEE Conference on Computer

Communications. Presented at the IEEE INFOCOM 2019 - IEEE Conference on Computer Communications, pp. 1495–1503. https://doi.org/10.1109/INFOCOM.2019.8737533
Haddaji, N., Bayati, A., Nguyen, K., Cheriet, M., 2018. BackHauling-as-a-Service (BHaaS) for 5G Optical Sliced Networks: An Optimized TCO Approach. Journal of Lightwave Technology 36, 4006–4017. https://doi.org/10.1109/JLT.2018.2855148
Hall, J.W., Thacker, S., Ives, M.C., Cao, Y., Chaudry, M., Blainey, S.P., Oughton, E.J., 2016a. Strategic analysis of the future of national infrastructure. Proceedings of the Institution of Civil Engineers - Civil Engineering 1–9. https://doi.org/10.1680/jcien.16.00018
Hall, J.W., Tran, M., Hickford, A.J., Nicholls, R.J., 2016b. The Future of National Infrastructure: A System-of-Systems Approach. Cambridge University Press.
Huang, J., Wang, C.-X., Feng, R., Sun, J., Zhang, W., Yang, Y., 2017. Multi-Frequency mmWave Massive MIMO Channel Measurements and Characterization for 5G Wireless Communication Systems. IEEE Journal on Selected Areas in Communications 35, 1591–1605. https://doi.org/10.1109/JSAC.2017.2699381
Intel, 2020. Comparing 5G vs. Wi-Fi 6 [WWW Document]. Intel. URL https://www.intel.com/content/www/us/en/wireless-network/5g-technology/5g-vs-wifi.html (accessed 12.27.20).
Jungnickel, V., Manolakis, K., Zirwas, W., Panzner, B., Braun, V., Lossow, M., Sternad, M., Apelfröjd, R., Svensson, T., 2014. The role of small cells, coordinated multipoint, and massive MIMO in 5G. IEEE Communications Magazine 52, 44–51. https://doi.org/10.1109/MCOM.2014.6815892
Katsaros, K., Dianati, M., 2017. A Conceptual 5G Vehicular Networking Architecture, in: Xiang, W., Zheng, K., Shen, X. (Sherman) (Eds.), 5G Mobile Communications. Springer International Publishing, Cham, pp. 595–623. https://doi.org/10.1007/978-3-319-34208-5_22
Khorov, E., Kiryanov, A., Lyakhov, A., Bianchi, G., 2019. A Tutorial on IEEE 802.11ax High Efficiency WLANs. IEEE Communications Surveys Tutorials 21, 197–216. https://doi.org/10.1109/COMST.2018.2871099
Knieps, G., Stocker, V. (Eds.), 2019. The Future of the Internet: Innovation, Integration and Sustainability, 1 edition. ed. Nomos Verlag.
Lehr, W., McKnight, L.W., 2003. Wireless Internet access: 3G vs. WiFi? Telecommunications Policy, Compeitition in Wireless: Spectrum, Service and Technology Wars 27, 351–370. https://doi.org/10.1016/S0308-5961(03)00004-1
Light Reading, 2019. Will 5G Kill WiFi? Qualcomm Thinks It Just Might [WWW Document]. Light Reading. URL https://www.lightreading.com/mobile/5g/will-5g-kill-wifi-qualcomm-thinks-it-just-might/d/d-id/749618 (accessed 7.21.20).
Lopez-Perez, D., Garcia-Rodriguez, A., Galati-Giordano, L., Kasslin, M., Doppler, K., 2019. IEEE 802.11be Extremely High Throughput: The Next Generation of Wi-Fi Technology Beyond 802.11ax. IEEE Communications Magazine 57, 113–119. https://doi.org/10.1109/MCOM.001.1900338
Mansell, R., 1999. Information and communication technologies for development: assessing the potential and the risks. Telecommunications policy 23, 35–50.
Massaro, M., 2017. Next generation of radio spectrum management: Licensed shared access for 5G. Telecommunications Policy, Optimising Spectrum Use 41, 422–433. https://doi.org/10.1016/j.telpol.2017.04.003
Massaro, M., Beltrán, F., 2020. Will 5G lead to more spectrum sharing? Discussing recent developments of the LSA and the CBRS spectrum sharing frameworks. Telecommunications Policy 44, 101973. https://doi.org/10.1016/j.telpol.2020.101973
Matinmikko, M., Latva-aho, M., Ahokangas, P., Seppänen, V., 2018. On regulations for 5G: Micro licensing for locally operated networks. Telecommunications Policy, The implications of 5G networks: Paving the way for mobile innovation? 42, 622–635. https://doi.org/10.1016/j.telpol.2017.09.004

Merlin, S., 2015. IEEE P802.11 Wireless LANs: TGax Simulation Scenarios (No. doc. : IEEE 802.11-14/0980r16). IEEE, New York.

Mir, Z.H., Filali, F., 2014. LTE and IEEE 802.11p for vehicular networking: a performance evaluation. J Wireless Com Network 2014, 89. https://doi.org/10.1186/1687-1499-2014-89

Mumtaz, S., Rodriguez, J., Dai, L., 2016. mmWave Massive MIMO: A Paradigm for 5G. Academic Press.

Navío-Marco, J., Arévalo-Aguirre, A., Pérez-Leal, R., 2019. WiFi4EU: Techno-economic analysis of a key European Commission initiative for public connectivity. Telecommunications Policy 43, 520–530. https://doi.org/10.1016/j.telpol.2018.12.008

Netflix, 2020. Internet Connection Speed Recommendations [WWW Document]. Help Center. URL https://help.netflix.com/en/node/306 (accessed 2.9.20).

Niu, Y., Li, Y., Jin, D., Su, L., Vasilakos, A.V., 2015. A survey of millimeter wave communications (mmWave) for 5G: opportunities and challenges. Wireless Netw 21, 2657–2676. https://doi.org/10.1007/s11276-015-0942-z

Ofcom, 2020. Ofcom Nations and Regions Technology Tracker - H1 2020 [WWW Document]. URL https://www.ofcom.org.uk/__data/assets/pdf_file/0037/194878/technology-tracker-2020-uk-data-tables.pdf (accessed 6.10.20).

Oughton, E.J., Frias, Z., 2016. Exploring the cost, coverage and rollout implications of 5G in Britain: A report for the UK's National Infrastructure Commission. Centre for Risk Studies, Cambridge Judge Business School, Cambridge.

Oughton, E.J., Frias, Z., Dohler, M., Crowcroft, J., Cleevely, D.D., Whalley, J., Sicker, D., Hall, J.W., 2017. The Strategic National Infrastructure Assessment of Digital Communications. Cambridge Judge Business School Working Paper 02/2017.

Oughton, E.J., Katsaros, K., Entezami, F., Kaleshi, D., Crowcroft, J., 2019. An Open-Source Techno-Economic Assessment Framework for 5G Deployment. IEEE Access 7, 155930–155940. https://doi.org/10.1109/ACCESS.2019.2949460

Oughton, E.J., Russell, T., 2020. The importance of spatio-temporal infrastructure assessment: Evidence for 5G from the Oxford–Cambridge Arc. Computers, Environment and Urban Systems 83, 101515. https://doi.org/10.1016/j.compenvurbsys.2020.101515

Panzner, B., Zirwas, W., Dierks, S., Lauridsen, M., Mogensen, P., Pajukoski, K., Miao, D., 2014. Deployment and implementation strategies for massive MIMO in 5G, in: 2014 IEEE Globecom Workshops (GC Wkshps). Presented at the 2014 IEEE Globecom Workshops (GC Wkshps), pp. 346–351. https://doi.org/10.1109/GLOCOMW.2014.7063455

Papadopoulos, H., Wang, C., Bursalioglu, O., Hou, X., Kishiyama, Y., 2016. Massive MIMO Technologies and Challenges towards 5G. IEICE TRANSACTIONS on Communications E99-B, 602–621.

Parker, M., Acland, A., Armstrong, H.J., Bellingham, J.R., Bland, J., Bodmer, H.C., Burall, S., Castell, S., Chilvers, J., Cleevely, D.D., Cope, D., Costanzo, L., Dolan, J.A., Doubleday, R., Feng, W.Y., Godfray, H.C.J., Good, D.A., Grant, J., Green, N., Groen, A.J., Guilliams, T.T., Gupta, S., Hall, A.C., Heathfield, A., Hotopp, U., Kass, G., Leeder, T., Lickorish, F.A., Lueshi, L.M., Magee, C., Mata, T., McBride, T., McCarthy, N., Mercer, A., Neilson, R., Ouchikh, J., Oughton, E.J., Oxenham, D., Pallett, H., Palmer, J., Patmore, J., Petts, J., Pinkerton, J., Ploszek, R., Pratt, A., Rocks, S.A., Stansfield, N., Surkovic, E., Tyler, C.P., Watkinson, A.R., Wentworth, J., Willis, R., Wollner, P.K.A., Worts, K., Sutherland, W.J., 2014. Identifying the Science and Technology Dimensions of Emerging Public Policy Issues through Horizon Scanning. PLoS ONE 9, e96480. https://doi.org/10.1371/journal.pone.0096480

Rangan, S., Rappaport, T.S., Erkip, E., 2014. Millimeter-wave cellular wireless networks: Potentials and challenges. Proceedings of the IEEE 102, 366–385.

Rendon Schneir, J., Ajibulu, A., Konstantinou, K., Bradford, J., Zimmermann, G., Droste, H., Canto, R., 2019. A business case for 5G mobile broadband in a dense urban area. Telecommunications Policy. https://doi.org/10.1016/j.telpol.2019.03.002

Roh, W., Seol, J.-Y., Park, J., Lee, B., Lee, J., Kim, Y., Cho, J., Cheun, K., Aryanfar, F., 2014. Millimeter-wave beamforming as an enabling technology for 5G cellular communications: theoretical feasibility and prototype results. IEEE Communications Magazine 52, 106–113. https://doi.org/10.1109/MCOM.2014.6736750

Roser, M., Ritchie, H., Ortiz-Ospina, E., 2019. Internet. Our World in Data.

Saint, M., Brown, T.X., 2019. A dynamic policy license for flexible spectrum management. Telecommunications Policy 43, 23–37. https://doi.org/10.1016/j.telpol.2018.07.002

Selinis, I., Filo, M., Vahid, S., Rodriguez, J., Tafazolli, R., 2016. Evaluation of the DSC algorithm and the BSS color scheme in dense cellular-like IEEE 802.11 ax deployments, in: 2016 IEEE 27th Annual International Symposium on Personal, Indoor, and Mobile Radio Communications (PIMRC). IEEE, pp. 1–7.

Sohul, M.M., Yao, M., Yang, T., Reed, J.H., 2015. Spectrum access system for the citizen broadband radio service. IEEE Communications Magazine 53, 18–25. https://doi.org/10.1109/MCOM.2015.7158261

Souryal, M.R., Nguyen, T.T., 2019. Effect of Federal Incumbent Activity on CBRS Commercial Service, in: 2019 IEEE International Symposium on Dynamic Spectrum Access Networks (DySPAN). Presented at the 2019 IEEE International Symposium on Dynamic Spectrum Access Networks (DySPAN), pp. 1–5. https://doi.org/10.1109/DySPAN.2019.8935639

Stocker, V., Smaragdakis, G., Lehr, W., Bauer, S., 2017. The growing complexity of content delivery networks: Challenges and implications for the Internet ecosystem. Telecommunications Policy, Celebrating 40 Years of Telecommunications Policy – A Retrospective and Prospective View 41, 1003–1016. https://doi.org/10.1016/j.telpol.2017.02.004

Taufique, A., Jaber, M., Imran, A., Dawy, Z., Yacoub, E., 2017. Planning Wireless Cellular Networks of Future: Outlook, Challenges and Opportunities. IEEE Access 5, 4821–4845. https://doi.org/10.1109/ACCESS.2017.2680318

TechRadar, 2019. 5G vs Wi-Fi 6 [WWW Document]. TechRadar. URL https://www.techradar.com/news/5g-vs-wi-fi-6 (accessed 12.27.20).

Vuojala, H., Mustonen, M., Chen, X., Kujanpää, K., Ruuska, P., Höyhtyä, M., Matinmikko-Blue, M., Kalliovaara, J., Talmola, P., Nyström, A.-G., 2019. Spectrum access options for vertical network service providers in 5G. Telecommunications Policy 101903. https://doi.org/10.1016/j.telpol.2019.101903

Wang, N., Hossain, E., Bhargava, V.K., 2015. Backhauling 5G small cells: A radio resource management perspective. IEEE Wireless Communications 22, 41–49. https://doi.org/10.1109/MWC.2015.7306536

WBA, NGMN Alliance, 2019. RAN Convergence Paper by WBA and NGMN Alliance. Wireless Broaband Alliance, Singapore.

Weiss, M.B.H., Werbach, K., Sicker, D.C., Bastidas, C.E.C., 2019. On the Application of Blockchains to Spectrum Management. IEEE Transactions on Cognitive Communications and Networking 5, 193–205. https://doi.org/10.1109/TCCN.2019.2914052

Weiss, T.A., Jondral, F.K., 2004. Spectrum pooling: an innovative strategy for the enhancement of spectrum efficiency. IEEE Communications Magazine 42, S8-14. https://doi.org/10.1109/MCOM.2004.1273768

West, J., Mace, M., 2010. Browsing as the killer app: Explaining the rapid success of Apple's iPhone. Telecommunications Policy 34, 270–286. https://doi.org/10.1016/j.telpol.2009.12.002

Wisely, D., Wang, N., Tafazolli, R., 2018. Capacity and costs for 5G networks in dense urban areas. IET Communications 12, 2502–2510. https://doi.org/10.1049/iet-com.2018.5505

Yaghoubi, F., Mahloo, M., Wosinska, L., Monti, P., de Souza Farias, F., Costa, J.C.W.A., Chen, J., 2018. A techno-economic framework for 5g transport networks. IEEE wireless communications 25, 56–63.

Yrjölä, S., Jette, A., 2019. Assessing the Feasibility of the Citizens Broadband Radio Service Concept for the Private Industrial Internet of Things Networks, in: Kliks, A., Kryszkiewicz, P., Bader, F., Triantafyllopoulou, D., Caicedo, C.E., Sezgin, A., Dimitriou, N., Sybis, M. (Eds.), Cognitive Radio-Oriented Wireless Networks, Lecture Notes of the Institute for Computer Sciences, Social Informatics and Telecommunications Engineering. Springer International Publishing, Cham, pp. 344–357. https://doi.org/10.1007/978-3-030-25748-4_26